# Tuning the water intrinsic permeability of PEGDA hydrogel membranes by adding free PEG chains of varying molar masses

Malak Alaa Eddine[1,2], Alain Carvalho[3], Marc Schmutz[3], Thomas Salez[4], Sixtine de Chateauneuf-Randon[1], Bruno Bresson[1], Nadège Pantoustier[1], Cécile Monteux*[1], Sabrina Belbekhouche*[2]

1- Laboratoire Sciences et Ingénierie de la Matière Molle, ESPCI Paris, 10 rue Vauquelin, Cedex 05 75231 Paris, France.
2- Université Paris Est Creteil, CNRS, Institut Chimie et Matériaux Paris Est, UMR 7182, 2 Rue Henri Dunant, 94320 Thiais, France.
3- Université de Strasbourg, CNRS, Institut Charles Sadron, 23 rue du Loess, 67034 Strasbourg Cedex 02, France.
4- Univ. Bordeaux, CNRS, LOMA, UMR 5798, F-33400 Talence, France.

* Authors for correspondence:
E-mail addresses: cecile.monteux@espci.fr (C. Monteux).

sabrina.belbekhouche@cnrs.fr (S. Belbekhouche).

**Abstract**

We explore the effect of poly (ethylene glycol) (PEG) molar mass on the intrinsic permeability and structural characteristics of poly (ethylene glycol) diacrylate PEGDA/PEG composite hydrogel membranes. We observe that by varying the PEG content and molar mass, we can finely adjust the water intrinsic permeability over several orders of magnitude. Notably, we show the existence of a maximum water intrinsic permeability, already identified in a previous study to be located at the critical overlap concentration $C^*$ of PEG chains, for the highest PEG molar mass studied. Furthermore, we note that the maximum intrinsic permeability follows a non-monotonic evolution with respect to the PEG molar mass and reaches its peak at 35 000 g.mol$^{-1}$. Besides our results show that a significant fraction of PEG chains is irreversibly trapped within the PEGDA matrix even for the shortest molar masses down to 600 g.mol$^{-1}$. This observation suggests the possibility of covalent grafting of PEG chains to the PEGDA matrix. CryoSEM and AFM measurements demonstrate the presence of large micron-sized cavities separated by PEGDA-rich walls whose nanometric structure strongly depends on the PEG content. By combining our permeability and structural measurements, we suggest that the PEG chains trapped inside the PEGDA rich walls induce nanoscale defects in the cross-linking density, resulting in an increased permeability below $C^*$. Conversely, above $C^*$, we speculate that partially-trapped PEG chains may form a brush-like arrangement on the surface of the PEGDA-rich walls, leading to a reduction in permeability. These two opposing effects are anticipated to exhibit molar-mass-dependent trends, contributing to the non-monotonic variation of the maximum intrinsic permeability at $C^*$. Overall, our results demonstrate the potential to fine-tune the properties of hydrogel membranes, offering new opportunities in separation applications.

**Introduction**

Macroporous polymer hydrogels have been the subject of an intense research, with applications in a wide range of areas [1], such as drug delivery [2-3], scaffold for tissue engineering [4], membrane for filtration processes [5] and environmental applications [6-7]. The porosity of the polymeric network can be obtained by various methods [8], such as foaming [9], emulsion templating [10], sacrificial templates such as salt crystals [11-13] or sacrificial particles such as calcium carbonate particles [14] or gelatin microparticles [15-18] and cryogelation [19]. While these methods involve several steps, phase separation is another simple approach to produce porous materials [20-23]. Polymerization Induced Phase Separation (PIPS) is a simpler method where phase separation occurs during the polymerization of an initially homogeneous solution of monomer and solvent [24-33] that is subsequently kinetically frozen. This method allows the production of hydrogel networks of varying morphology, mechanical properties, and intrinsic permeability[34]. The intrinsic permeability is a measure of the relative ease with which a porous medium can transmit a fluid under a potential gradient and is a property of the medium alone [35].

PEGDA-based hydrogels have been shown to undergo PIPS which generates water cavities in the PEGDA network. Alternatively, adding incompatible polymers in the PEGDA solution such as polyurethane (PU), polyetherimide (PEI) or hyaluronic acid (HA) [36-38] also enables the formation of a porous structure. PEGDA-based membranes are promising for filtration, gas separation, tissue engineering, or drug-delivery applications, because they are hydrophilic, resistant and their macroporous structure enables a good control of their transport properties with respect to water, gas, or nutrients [39-40].

We have shown in previous works[41-42] that adding long PEG chains in the PEGDA prepolymerization solution enables to obtain macroporous hydrogels with a controlled permeability varying by orders of magnitude with the PEG concentration. We found that the resulting hydrogels present micron-size cavities separated by PEGDA-rich walls. Indeed, the intrinsic permeability of these composite PEG/PEGDA hydrogel systems can be varied by orders of magnitude by tuning the PEG content. We also showed that these long PEG chains remain irreversibly trapped in the hydrogels and are not rinsed out – even after several filtration cycles and pressures up to 100 kPa. Hence, the free PEG chains are not porogens and the induced macroporosity cannot be attributed to the rinsing of the chains after polymerization of the PEGDA network, which is in contradiction with several studies of the literature [43-46]. It should be noted that, to our knowledge, it has never been demonstrated that PEG was actually removed in these studies, whereas in our study we are directly investigating its presence or not. Besides, in our latest work [42], we further showed that 20-nm nanoparticles permeate through the most permeable PEG/PEGDA hydrogel, while the 100-nm and 1-µm particles do not, suggesting that the PEG chains trapped in the PEGDA matrix induce local nanometric defects in the cross-linking density controlling the permeability. This effect should in principle depend on the molar mass of the PEG chains. In this study, our goal is to determine how the molar mass of the PEG chains influences the intrinsic permeability and structure of the PEG/PEGDA hydrogels. We demonstrate that the intrinsic permeability maximum obtained at $C = C^*$ is a general and robust feature observed for the various PEG chains investigated.

**Experimental section**

**Materials**

Poly (ethylene glycol) diacrylate PEGDA ($\overline{Mw}$ = 700 g.mol$^{-1}$) with 13 ethylene oxide units and 4-(2-hydroxyethoxy) phenyl 2-hydroxy-2-propyl ketone (Irgacure 2959) photoinitiator were purchased from Sigma–Aldrich. Linear poly (ethylene glycol)s (PEG) ($\overline{Mw}$ = 600 ; 3000 (polydispersity index Đ = $\overline{Mw}/\overline{Mn}$ = 1.031) ; 10 000 (Đ = 1.031) ; 35 000 (Đ = 1.026)) were purchased from Sigma–Aldrich. Linear poly (ethylene glycol)s (PEG) ($\overline{Mw}$ = 300 000 g.mol$^-$ [1], Đ = 2.1) and ($\overline{Mw}$ = 600 000 g.mol$^{-1}$) were purchased from Serva and Acros Organics, respectively. Water was purified with a Milli-Q reagent system (Millipore).

**PEGDA hydrogels preparation**

The PEGDA and PEG/PEGDA membranes were synthesized via UV-initiated free-radical photopolymerization using Irgacure 2959 as the photoinitiator, as described in our previous studies [41-42]. The exact compositions are given in the Supporting Information (see Tab. S1-S6). Briefly, prepolymerization solutions are composed of a fixed mass of PEGDA (2 g), Irgacure (2 mg), and water ( 10 g) and increasing amounts of PEGs (between 0.01 g and 5 g), depending on the PEG molar mass. We chose this recipe to keep the PEGDA/water ratio constant and equal to 0.2, as this parameter was demonstrated to play an important role in PEGDA hydrogel structure [47]. Moreover, the Irgacure/PEGDA ratio was fixed to 0.1 wt% which is similar to several values from the literature [48]. The wt% of PEG varies between 0.08 and 29.5 wt%, depending on the PEG molar mass. In the Supporting Information (see Tab. S1-S6) all the compositions used are provided. To prepare the hydrogel membranes, the prepolymerization solutions were sandwiched between two glass plates (120 mm x 80 mm) which were separated by 1 mm thick spacers leading to membranes with 1 mm thickness. Then, the solution was polymerized under irradiation of UV light (intensity =1800 μw/cm$^2$) with a wavelength of 365 nm for 10 min, similarly to other studies [48-49]. After polymerization, the obtained hydrogel sheets were placed in a petri dish with pure water for at least 24 hours prior filtration to eliminate any unreacted PEGDA monomers or remaining free PEG chains. The fraction of PEG washed out of the gel in the supernatant or during filtration experiments was determined using Total Organic Carbon (TOC) measurements. To prepare the hydrogel membranes, these large hydrogel sheets were then cut with a round die puncher of 45 mm in diameter to obtain 1-mm thick disks with 45 mm diameter.

In order to confirm that the addition of PEG with different molar masses does not affect the polymerization reaction of PEGDA, as in our previous work [41], we follow the polymerization reaction of PEGDA hydrogels prepared with different concentrations of PEG-3000 g.mol$^{-1}$. We obtain that the PEGDA polymerization was complete even after the addition of PEG chains with a concentration ranging between 0 and 20 wt%. The results are presented in the Supporting Information (see Fig. S1). Moreover, we have evidenced also that the water content is between 80 and 82% for all the hydrogels (PEGDA/water and PEGDA/water +

PEG) and that no significant swelling is measured when the hydrogels are immersed in pure water at room temperature [41].

**Morphological characterization of hydrogel membranes**

*Atomic Force Microscopy (AFM)*

AFM images were obtained with a Bruker Icon microscope driven by a Nanoscope V controller. The surface of the hydrogel membrane immersed in water before filtration was observed in Peak Force mode. The height images were acquired with a cantilever of spring constant 0.7 N.m$^{-1}$ specially designed for this application. In this mode, similar to a rapid approach-retract experiment, the cantilever oscillates at a frequency of 1 kHz. The scanning frequency was 0.7 Hz and the maximum force was set to 500 pN. We chose to use AFM to image the membrane surface in situ in water.

*CryoScanning Electron Microscopy (CryoSEM)*

PEGDA and PEGDA/PEG hydrogel membranes with a thickness of 1 mm were placed on a homemade cryo-holder to be quickly plunged into an ethane slush. As the sample was free-standing over the holder, the sample was rapidly frozen during the plunging by direct contact with the liquid ethane, in order to form amorphous ice. Subsequently, the sample was transferred into the Quorum PT 3010 chamber attached to the microscope. There, the frozen sample was fractured with a razor blade. A slight etching at - 90 °C may be performed to render the sample more visible. The sample was eventually transferred in the FEG-cryo SEM (Hitachi SU8010) and observed at 1 kV at −150 °C. No further metallization step was required before transferring the sample to the SEM chamber. Several sublimation cycles were performed on each sample to ensure the removal of the ice from the hydrogel.

**Filtration experiments**

Water intrinsic permeability through PEGDA and PEG-modified PEGDA hydrogels was measured using a dead-end ultrafiltration UF cell obtained from Fisher Scientific S.A.S. (Amicon Model 8050, 50 mL for 45 mm Filters) as described in our previous study [41] (see Supporting Information Fig.S2). The filtration was performed at ambient temperature, with Milli-Q water as the feed solution. The membrane with an area of 15.90 cm$^2$ was fixed in the membrane holder of the cell. Standard filtration experiments were performed at a pressure

P=10 kPa and the liquid permeate was weighted as a function of time with a balance to obtain a precise measurement of the flow rate $Q$.

The permeate flow rate $Q$ was recorded, and the water intrinsic permeability $K$ was calculated from the following relation:

$$K = \frac{Q\mu h}{\Delta P S} \qquad , \qquad (1)$$

where $Q$ is the water flow rate calculated from the slope of the variation of the accumulated permeate volume as a function of time, $\mu$ is the water viscosity, $h$ is the hydrogel thickness, $S$ is the surface area of the hydrogel membrane and $\Delta P$ is the pressure difference across the membrane. In this study, the values of intrinsic permeability $K$ that are presented are obtained from the slopes of the $Q = f(\Delta P)$ curves, for pressures ranging from 0 to 10 kPa where the flux varies linearly with the applied pressure (Fig.S3).

**Results**

*Permeability variations around $C^*$ for varying PEG molar masses*

To obtain the water intrinsic permeability $K$ of the PEG/PEGDA hydrogels, we measure the water flux $Q$ as a function of the applied pressure, for pressures such that $P < 10$ kPa in the linear regime for which $Q$ varies linearly with $\Delta P$. In Figure 1, we report the intrinsic permeability measurements for a series of PEGDA/PEG hydrogel systems with the molar mass of PEG ranging from 600 to 600 000 g.mol$^{-1}$. For molar masses between 600 and 3000 g.mol$^{-1}$, the water intrinsic permeability continuously increases with the PEG content. As an example, for PEGDA prepared with PEG-600 g.mol$^{-1}$ (Figure 1 a), the water intrinsic permeability varies over nearly two orders of magnitude from $3.6 \pm 0.0015 \times 10^{-18}$ m$^2$ to $0.8 \pm 0.05 \times 10^{-16}$ m$^2$ when the PEG content increases from 7.7 to 29.5 wt%. In comparison, the water intrinsic permeability for PEGDA/PEG-3000 g.mol$^{-1}$ hydrogels varies by about three orders of magnitude with the PEG content, from $4.6 \pm 0.0025 \times 10^{-18}$ m$^2$ to $8.5 \pm 1.6 \times 10^{-16}$ m$^2$ when the PEG content increases from 1.6 to 20 wt% (Figure 1 b). For the highest PEG molar mass (10 000, 35 000, 300 000 and 600 000 g.mol$^{-1}$), the water intrinsic permeability presents a maximum with the PEG concentration. For PEG-300 000 g.mol$^{-1}$, we have already shown that this maximum is obtained for $C = C^* = 1.6$ wt%, i.e. the overlap concentration of PEG chains which marks the transition between the semi-dilute and concentrated regimes [41]. For the other molar masses, $\overline{Mw} = 10\ 000$ g.mol$^{-1}$, $35\ 000$ g.mol$^{-1}$ and $600\ 000$ g.mol$^{-1}$, the intrinsic permeability maximum are respectively obtained for $C \approx 8$, 4 and 1 wt% which closely correspond to the values of $C^*$ presented in Table 1. These values of $C^*$ are obtained either by a change of slope in the viscosity measurements $\mu(C)$ (see Supporting Information, Fig. S4) or by using the following polymer physics scaling law:

$$C^* = \frac{\overline{Mw}}{\frac{4}{3}\pi r_g^3 N_A}, \qquad (2)$$

where $\overline{Mw}$ is the weight molar mass of the polymer, $r_g$ is the gyration radius of polymer coils and $N_A = 6.023 \times 10^{23}$ mol$^{-1}$ is Avogadro's number. The gyration radius $r_g$ is calculated from the chain end-to-end distance $R$, in dilute conditions, by:

$$r_g = \frac{R^{1/2}}{\sqrt{6}} = \frac{(C_\infty n l^2)^{\frac{1}{2}}}{\sqrt{6}}, \qquad (3)$$

where $l$ is the length of the C-C bond ( ~1.54 Å), $n$ is the number of ethylene glycol monomers (with $M_w$=44 g.mol$^{-1}$) in a polymer chain, and $C_\infty$ is Flory's characteristic ratio for PEG in water which increases with the chain length [50]. For the highest molar masses, we take $C_\infty \sim 5$ consistently with previous work [51]. For the lowest PEG molar masses, we take $C_\infty \sim 2.7$.

*Table 1. Values of $r_g$ and $C^*$ for different PEG molar masses.*

| PEG molar mass (g.mol$^{-1}$) | 600 | 3000 | 10 000 | 35 000 | 300 000 | 600 000 |
|---|---|---|---|---|---|---|
| $r_g$ (nm) | 0.7 | 1.6 | 3.7 | 6.8 | 20.1 | 28.4 |
| $C^*$ (wt%) | above 50 | ~37 | 8 | 4.2 | 1.6 | 1 |

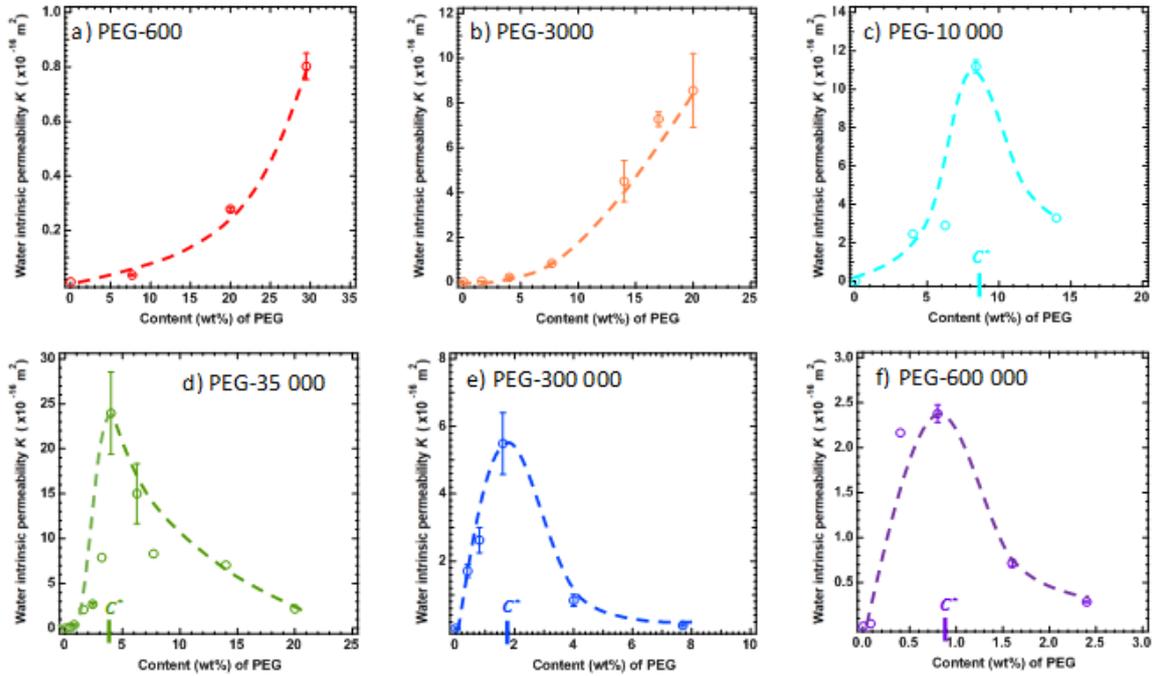

*Figure 1. Water intrinsic permeability at pressure P=10 kPa versus content of a) PEG-600 g.mol$^{-1}$; b) PEG-3000 g.mol$^{-1}$; c) PEG-10 000 g.mol$^{-1}$; d) PEG-35 000 g.mol$^{-1}$; e) PEG-300 000 g.mol$^{-1}$ and f) PEG-600 000 g.mol$^{-1}$ in the prepolymerization mixture of hydrogel membranes prepared with 16 wt% of PEGDA. The dashed lines are guides for the eye.*

In order to characterize in more details the dependence of the hydrogel permeability on the PEG molar mass, we plot, on Figure 2 a, the water intrinsic permeability $K$ as a function of the PEG content $C$ (wt%) for all PEG molar mass. At low PEG content, we observe that the slope increases when the PEG molar mass increases. Moreover, Figure 2 b shows that the maximal water intrinsic permeability $K^*$ – obtained at the critical overlap concentration of PEG chains – presents a maximum for a PEG molar mass of 35 000 g.mol$^{-1}$. The variation of $K^*$ is non monotonic, hence these two results prevent the possibility of rescaling easily the $K$ ($C$) curves onto a single master curve. It deserves noting that we have tried to rescale the curves by plotting $K/K^*$ as a function of $C/C^*$ but the curves cannot be rescaled satisfactorily (Fig. S5). The reason why the slope increases below $C^*$ it is believed that it is due to the size of the PEG chains that controls the size of the defects in the PEGDA matrix. Above $C^*$ it is probably a viscosifying effect that increases with molar mass hence leads to a more pronounced decrease of $K$ with molar mass.

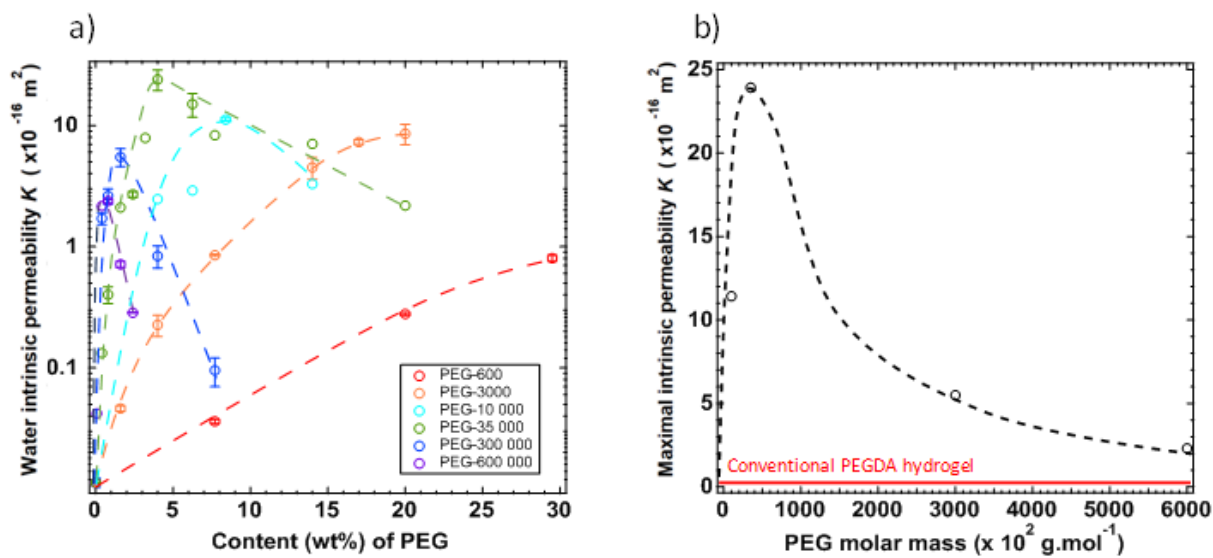

*Figure 2. a) Water intrinsic permeability as a function of PEG concentration for various molar masses. b) Maximal water intrinsic permeability as a function of PEG molar mass. The red solid line represents the water intrinsic permeability of the conventional PEGDA hydrogel membrane without PEG chains (K~ $10^{-18}$ $m^2$). The dashed lines are guides for the eye.*

We have summarized the most relevant findings in terms of water intrinsic permeability as a function of the PEG concentration conditions in Table 2.

*Table 2. Values of $K^*$ and $C^*_{exp}$ for different PEG molar masses.*

| PEG molar mass (g.mol$^{-1}$) | 600 | 3000 | 10 000 | 35 000 | 300 000 | 600 000 |
|---|---|---|---|---|---|---|
| $C^*_{exp}$ (wt%) | – | – | 8 | 4 | 1.6 | 1 |
| $K^*$ (x $10^{-16} m^2$) | 0.8 | 8.5 | 11.2 | 23.9 | 5.5 | 2.4 |

*Fraction of the PEG chains trapped in the PEGDA matrix*

In this section, our goal is to determine whether the molar mass of the PEG chains influences the amount of chains trapped inside the resulting hydrogel. We analyze the content of PEG released from the hydrogel in the water supernatant, where it is kept overnight, as well as in the permeate after a cyclic filtration, by total organic carbon (TOC) analysis. During cyclic filtration, pressure is systematically raised from 0 to 1 Bar, followed by a gradual decrease, repeating this process three times to enhance purification efficiency. Depending on the content and the molar mass of PEG, the three cycles of filtration can take anywhere from 1 to several hours to complete. The fraction of PEG that we calculate is the ratio of PEG molecules found in the permeate divided by the total amount of PEG initially present in the hydrogels and is therefore calculated using the relation:

$$f = \frac{C_{permeate} * V_{permeate}}{\emptyset_{PEG} * \rho_{hydrogel} * V_{hydrogel}} \quad , \quad (4)$$

where $C_{permeate}$ is the concentration in the permeate measured by TOC, $V_{permeate}$ is the volume of permeate, $\emptyset_{PEG}$ is the weight fraction of PEG in the hydrogel, $\rho_{hydrogel}$ is the volumic mass of the hydrogel taken equal to $\rho_{water} = 10^6$ mg.L$^{-1}$ and $V_{hydrogel}$ is the volume of the hydrogel membrane.

As presented in Figure 3, the fraction of PEG calculated in the supernatant depends on the PEG molar mass. For the small PEGs (i.e. PEG-600 and 3000 g.mol$^{-1}$), 20% of the PEG initially present in the PEGDA/PEG hydrogel is released from the hydrogel membrane into the supernatant (Figures 3 a and b) while 80% remains trapped in the sample. However, less than 4% of the PEG chains can be found in the supernatant for the large PEG chains of 300 000 g.mol$^{-1}$ (Figure 3 c). In contrast, the fraction of PEG calculated in the permeate after a cyclic filtration is negligible (< 1%), regardless of the molar mass.

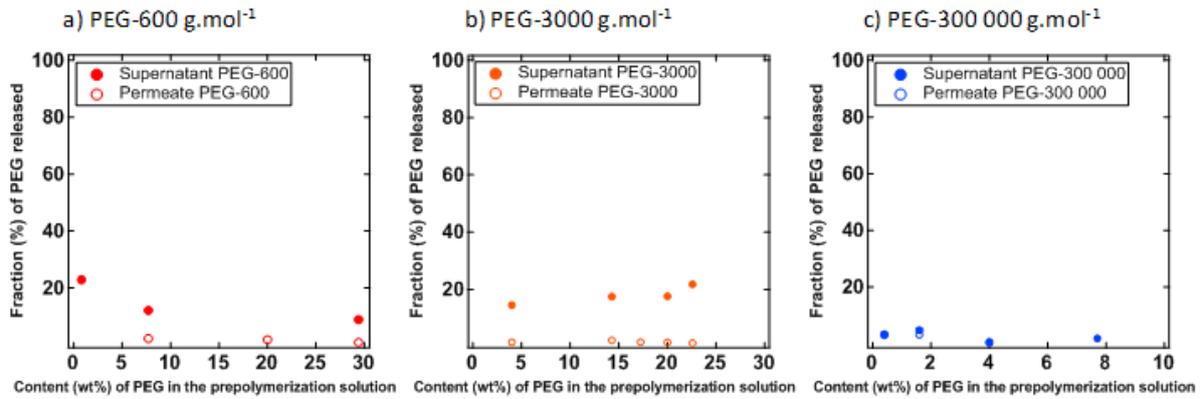

*Figure 3. Variation of the fraction of PEG in the supernatant liquid (solid symbols) and in the permeate after cyclic filtration (empty symbols) as a function of the content of PEG in the prepolymerization solution for a) PEG-600, b) PEG-3000 and c) PEG-300 000 g.mol$^{-1}$.*

To account for the irreversible incorporation of most PEG chains in the PEGDA matrix, a possible scenario is that the PEG chains are entropically trapped into the tight PEGDA matrix. This concept of entropic trapping was evidenced experimentally by Liu *et al.* [52] for Poly(styrene sulfonate) (PSS) chains trapped in a polyacrylamide hydrogel. Entropic trapping is expected to retard the diffusion of free chains in the tight hydrogels as their permeation would require a large deformation of the PEG chains which is unfavorable from an entropic point of view. Entropic trapping is supposed to occur for chains of intermediate size. Indeed, small chains are expected to diffuse like spheres while larger chains are expected to diffuse in a reptation-like manner. In our case, the chains seem to be partially trapped independently of their molar mass. We can suggest an alternative reason for the observed trapping of the chains. The PEG chains may be chemically grafted to the PEGDA matrix through a chain-transfer mechanism where the radicals transfer from the active PEGDA chains to the PEG chains. Such a chain-transfer mechanism has been observed by Nandi *et al.* [53] during the radical polymerization of methyl methacrylate (MMA) in the presence of various glycols through the rupture of the C-H bonds of the glycol next to the -OH group.

*Roles of PEG concentration and molar mass on the hydrogel structure*

In order to understand the relation between the PEGDA/PEG hydrogel permeability and structure and how this depends on the PEG content, we perform AFM and cryoSEM measurements. CryoSEM measurements are performed with PEG molar masses of 600 g.mol$^{-1}$, 3000 g.mol$^{-1}$, 300 000 g.mol$^{-1}$ and 600 000 g.mol$^{-1}$. However, for the sake of clarity, we only present in Figure 4 the results obtained with PEG 3000 g.mol$^{-1}$ and 600 000 g.mol$^{-1}$, while the other molar masses are presented in the Supporting Information (Fig. S6 and Fig. S7). As a reference, the case of PEGDA without PEG is given in Figure 4 a.

As discussed in Molina's article [47] and in our previous studies [41-42], the 200-nm large cavities observed with the PEGDA sample correspond to the water in excess that is rejected from the PEGDA matrix, whose equilibrium polymer/water fraction is 50/50 wt%/wt%. Given the low intrinsic permeability of the PEDGA samples, we suggest that these cavities are closed. Adding 4 wt% of PEG-3000 g.mol$^{-1}$ induces a slight increase in the size of the cavities up to ~ 500 nm (Figure 4 b). For some of the cavities, darker zones – marked with arrows – reveal areas which extend in depth in the direction perpendicular to the plane of observation and correspond to water that is sublimated during the successive sublimation cycles. These observations show that a weak fraction of the pores seems to be open in contrast to the case of pure PEGDA for which these cavities are closed. For the 17 wt% of PEG samples (Figure 4 c), the size of the cavities increases and reaches typically one micron. In this case also, some of the cavities seem to extend in depth.

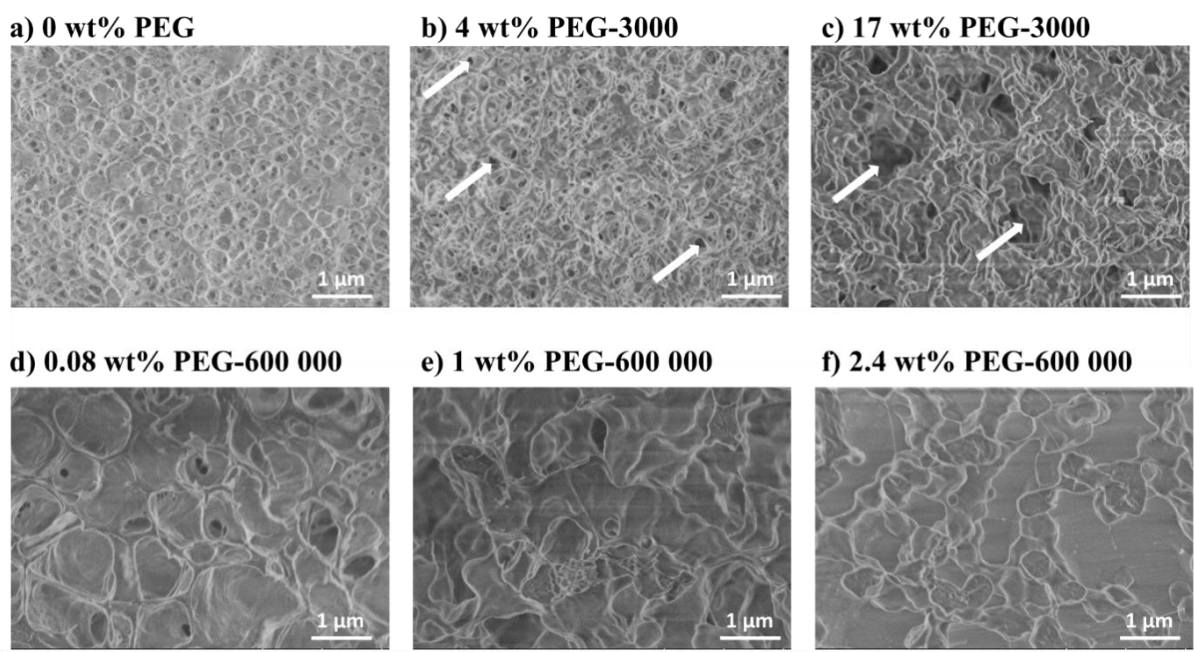

*Figure 4. CryoSEM images of hydrogel membranes prepared with PEGDA and various contents of PEG-3000 g.mol$^{-1}$: a) 0, b) 4 (before $C^*$) and c) 17 wt% (after $C^*$); as well as PEG-600 000 g.mol$^{-1}$: d) 0.08 (before $C^*$), e) 1 (at $C^*$) and f) 2.4 wt% (after $C^*$).*

For PEG-600 000 g.mol$^{-1}$, micron-sized cavities are observed similarly to the case of PEG-300 000 g.mol$^{-1}$ [41]. In this case, we do not detect any extension of the cavities in the direction perpendicular to the plane of observation. These results confirm previous filtration and permeation measurements from which we concluded that the micron-sized cavities probably do not form a percolating network and that the permeation of water and nanoparticles is rather controlled by the structure of the PEGDA-rich walls between these micron-sized cavities. To obtain more insight into the wall structure, we perform AFM measurements with a field of view of 10 μm for the various PEG molar masses at various concentrations below and above $C^*$. For reasons of image clarity, the z-scales are not identical for all images and we have chosen to adapt the z-scale at best for each image.

The AFM images obtained for PEGDA/PEG hydrogels prepared with PEG-3000 g.mol$^{-1}$ are represented in Figure 5. As a reference, the case of PEGDA without PEG is given in Figure 5 a. For 4 wt% of PEG added, we observe an increased number of the large cavities (Figure 5 b). The same trend is observed for PEG 600 and 10 000 g.mol$^{-1}$, as represented in the Supporting Information (see Fig.S7).

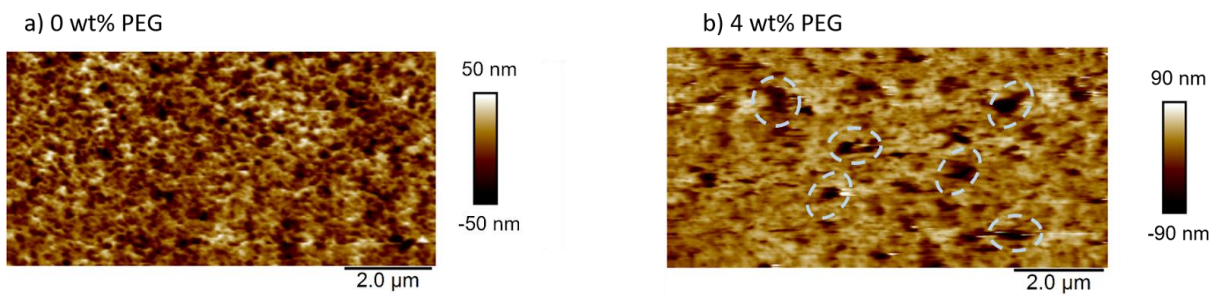

*Figure 5. Surface AFM images of PEGDA hydrogels membranes prepared with 16 wt % of PEGDA and various content of PEG-3000 g.mol$^{-1}$ a) 0 wt% and b) 4 wt%.*

For hydrogel prepared with PEG-35 000 g.mol$^{-1}$, a low content of PEG (i.e. 0.4 wt %) does not significantly modify the structure (Figure 6 a). When the PEG content increases to 0.8 wt%, some large cavities of size ~500 nm to 1 μm can be seen as well as small cavities of the order of tens of nanometers. When the PEG content increases to 2.4 wt% and then 4 wt% ($C^*$), micron size valleys and hills can be observed, which presumably correspond to the micron size cavities observed with cryoSEM. This micron size structure presents a very low roughness at the nanometric scale, similarly to the results obtained with PEG-300 000 g.mol$^{-1}$ and presented in the previous studies [41-42].

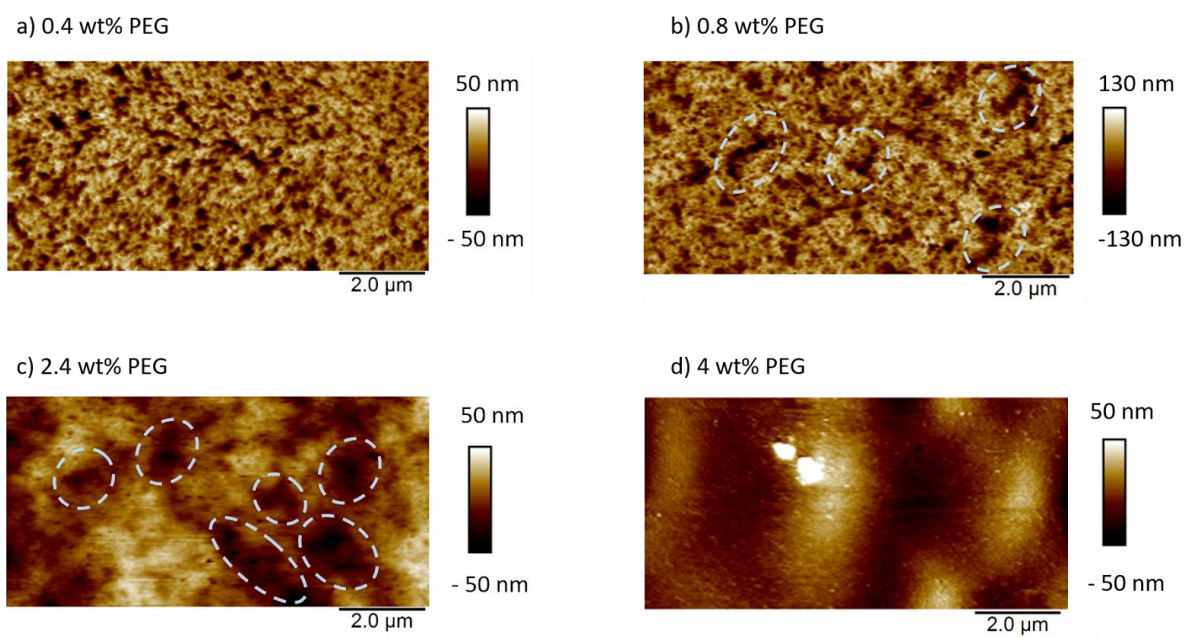

*Figure 6. Surface AFM images of PEGDA hydrogel membranes prepared with 16 wt % of PEGDA and various contents of PEG-35 000 g.mol$^{-1}$: a) 0.4 wt%, b) 0.8 wt%, c) 2.4 wt% and d) 4 wt%.*

**Discussion**

We observe a maximum of the intrinsic permeability with the PEG concentration being at $C^*$ except for the molar masses for which $C^*$ cannot be reached experimentally. The size of the cavities is ~200 nm for PEGDA samples while for all samples containing PEG the size of the cavities is larger and on the order of 1 μm, independently of the molar mass. The PEG chains are supposed to be trapped in the PEGDA matrix during the polymerization and remain trapped even after several cycles of permeation. We also recall our previous study showing for hydrogels below $C^*$ that 20-nm particles can permeate through the hydrogels while 100-nm particles cannot.

Combining all these results, we propose an evolution of the structure of PEGDA hydrogel with the PEG concentration, which is schematically illustrated in Figure 7. At PEG concentrations below $C^*$, we suggest that the PEG chains are trapped in the PEGDA-rich walls separating the micron-sized heterogeneities and provide local defects in the cross-linking density which enable an increase of the water permeation flux. Some PEG chains may also be trapped inside the water cavities. More precisely, the limiting factor of the pressure driven water transport is due to the presence of pores induced by the PEG chains in the PEGDA walls. The size of those pores depends on the molar mass of the PEG. Indeed, the permeability of a PEGDA membrane (20 wt% PEGDA - 80 wt% water) is very low and the 200 nm cavities are closed. For low PEG concentration (Figure 7 a), adding PEG will connect the cavities and increase the permeability. Above $C^*$, the PEG chains that are trapped in the cavities may entangle and form a dense network that controls the water intrinsic permeability which drops when the concentration further increases. It can be noted on Figure 7 that the typical size of the cavities is 1 micron while the size of PEG chains ranges from 1 nm to 30 nm. Let us now discuss some possible mechanisms for the evolution of the hydrogel structure. Molina *et al.*[47] have studied the structure of PEGDA/water samples (with no PEG) using Small-angle neutron scattering (SANS). According to their study, the equilibrium water volume fraction that the cross-linked PEGDA network can incorporate is 50 wt%, as a result of the balance between the osmotic pressure and the elasticity of the network. When the prepolymerization solution contains more than 50 wt% water, the water in the excess phase separates and forms water cavities of 200 nm in size that are probably generated during the fast UV polymerization process. The size of the cavities probably results from a complex interplay between polymerization rate, flow, and diffusion of water between the polymerized and non-polymerized zones. We suggest that the presence of the PEG chains may facilitate the coalescence of the water cavities during the polymerization process leading to the large 1-µm cavities, instead of the 200-nm ones in the absence of PEG. As discussed earlier, these PEG chains may either be covalently grafted to the PEGDA network through a radical-transfer mechanism or trapped for entropic reasons inside the tight PEGDA network. As the size of the cavities increases, the number of cavities and walls probably decreases, hence the number of trapped chains per wall increases, leading to a fast increase of the permeability with the PEG concentration. To account for the decrease of $K_{max}$ for molar masses above 35000 g.mol$^{-1}$ (Figure 2 b), we suggest that for the largest molar masses, the clogging of the cavities with the PEG chains may start at low concentration and may compensate the increase of permeability caused by the PEG chains trapped in the walls.

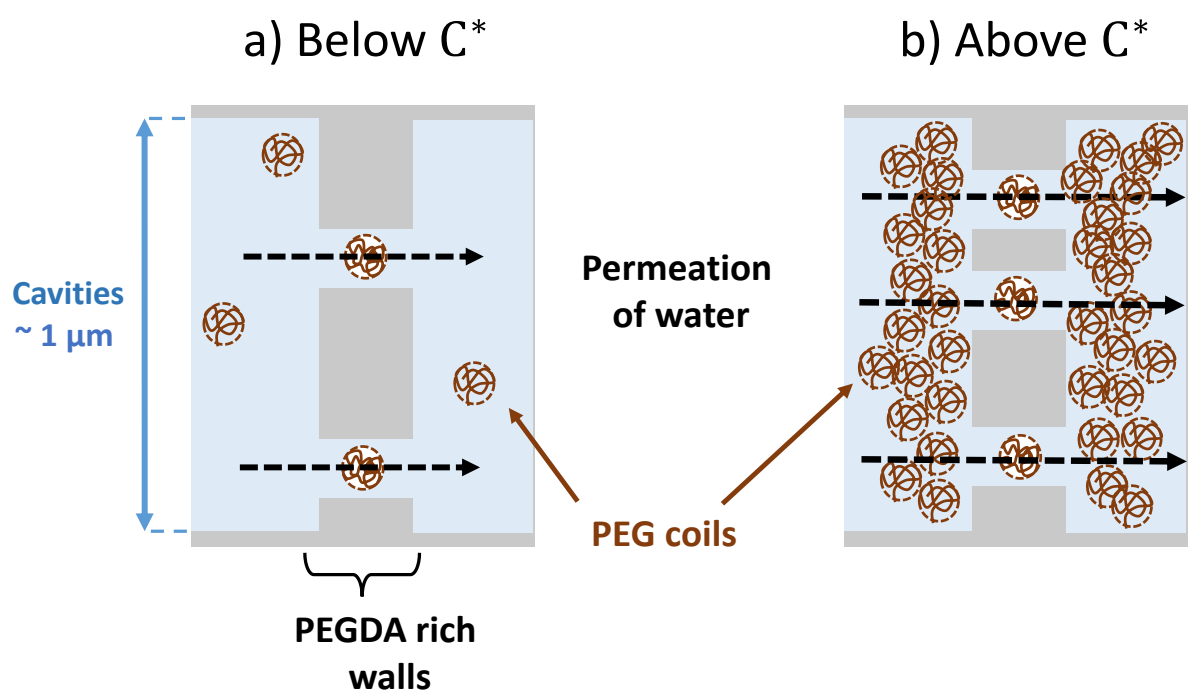

*Figure 7. Schematic of the proposed hydrogel structures below and above the critical concentration of PEG. a) At low PEG concentration, the PEG chains trapped in the walls separating the micron-sized cavities control the permeation of particles and water through the hydrogel. b) At higher PEG concentration, the PEG chains entangle together inside the water cavities, reducing the water intrinsic permeability.*

**Conclusion**

We developed a series of composite hydrogel membranes obtained by a simple photopolymerization of PEGDA under UV light in the presence of PEG chains with a wide range of molar masses. Compared to the studies from the literature, we have shown that regardless of the molar mass, most PEG chains remain irreversibly trapped in the hydrogel matrix during filtration experiments. Increasing the content of PEG chains in the hydrogel allowed us to tune the PEGDA/PEG water intrinsic permeability over several orders of magnitude. In addition, we showed that the existence of a maximum of the intrinsic permeability obtained for the critical overlap concentration of PEG is a robust phenomenon observed for a wide range of PEG molar masses. We show that the maximum intrinsic permeability at $C^*$ follows a non-monotonic trend with the molar mass and is maximum for PEG-35 000 g.mol$^{-1}$. We suggest that the permeability increase below $C^*$ is due to the PEG chains trapped in the PEGDA matrix that induce defects in the cross-linking density of the PEGDA walls between the micron-sized cavities observed with cryoSEM. Above $C^*$, the permeability decrease is suggested to be due to the fact that the PEG chains may form an entangled network inside the water cavities that reduces the permeability of the whole system. One of the possible applications of the designed hydrogel is the filtration of particles [42] depending on their size and we aim at extending this for environmental application, e.g. for the microplastic filtration.

**Acknowledgments**

We gratefully acknowledge Institut Carnot for microfluidics for the financial support during this research project, as well as the Agence Nationale de la Recherche (grants ANR-21-ERCC-0010-01 *EMetBrown,* ANR-21-CE06-0029 *Softer,* ANR-21-CE06-0039 *Fricolas*), and the European Union through the European Research Council (grant ERCCoG-101039103 *EMetBrown*). We also thank the Soft Matter Collaborative Research Unit, Frontier Research Center for Advanced Material and Life Science, Faculty of Advanced Life Science at Hokkaido University, Sapporo, Japan.